\documentclass[conference]{IEEEtran}
\IEEEoverridecommandlockouts
\usepackage{cite}
\usepackage{amsmath,amssymb,amsfonts}
\usepackage{algorithmic}
\usepackage{graphicx}
\usepackage{textcomp}
\usepackage{url}
\usepackage{multirow}
\usepackage{siunitx}
\def\BibTeX{{\rm B\kern-.05em{\sc i\kern-.025em b}\kern-.08em
    T\kern-.1667em\lower.7ex\hbox{E}\kern-.125emX}}
\usepackage{geometry}
\begin{document}

\title{A Customized System to Assess Foot Plantar Pressure: A Case Study on Calloused and Normal Feet\\
\thanks{This work was supported by the National Science Foundation, Sri Lanka grant RPHS-2016-DTM02 and the Senate Research Council of University of Moratuwa grant SRC/CAP/15/01.}
}

\author{\IEEEauthorblockN{$^{1}$A.M.A.I. Rathnayaka, $^{1}$$^{\dagger}$W.N.D. Perera, $^{1}$H.P. Savindu, $^{1}$K.C.M. Madarasingha, $^{1}$S.P. Ranasinghe, \\ $^{1}$H.G.T.V. Thuduwage, $^{1}$A.U. Kulathilaka, $^{1}$P. Silva, $^{2}$S. Jayasinghe, \\ $^{3}$K.T.D. Kahaduwa, , $^{1}$A.C. De Silva}
\IEEEauthorblockA{\textit{$^{1}$University of Moratuwa, Sri Lanka $^{2}$University of Colombo, Sri  Lanka $^{3}$National Hospital of Sri Lanka}\\
$^{\dagger}$130441f@uom.lk}
}

\maketitle

\begin{abstract}
Foot plantar pressure monitoring is an important tool for biomechanical assessment of posture, foot complications due to callus formation and wounds and for sports applications. The pronounced cost associated with commercial plantar pressure monitoring systems and inflexibility of custom analyzing data in such systems prompted the development of a versatile system with minimized cost. This study focuses on the development of such a system with high speed data acquisition providing analysis tools for assessing plantar pressure variations of diabetic patients with calloused feet. 
The new system is capable of achieving a frame rate of 155 Hz which is ideal for pressure monitoring during both standing and walking. The system was verified using 10 normal subjects and 5 diabetic subjects with calluses on in their feet. Results indicate significantly high mechanical stresses on skin beneath callus and postural disorders during standing, in subjects with calluses.  
\end{abstract}


\section{Introduction}
Excessive mechanical stress on the plantar foot is a major risk factor in forming ulcers and related foot complications which can be detected at early stages by assessing plantar foot pressure variations \cite{1_rai2006study}. 
%
%

One of the commonplace foot complications is diabetes related foot complications. All types of diabetes complications include peripheral vascular disease, peripheral neuropathy, limited joint mobility and bone deformity \cite{4_botros2016prediction}. The lack of sense in lower and upper limb extremities is the main characteristics of the above mentioned diseases. As a result, a higher mechanical stress is applied on the foot plantar without the knowledge of the patient. This is an important precursors to form callus and foot ulcers. Hence, it can be concluded that plantar pressure variations of the foot and the development of diabetes related foot ulcers are highly correlated.
It was reported that areas of callus formation of the plantar foot can increase the pressure up to 30\% \cite{1_rai2006study}, most commonly, areas under $1^{st}$, $2^{nd}$ and $5^{th}$ metatarsal heads and the heel \cite{7_amemiya2016establishment}.

The aim of this research was to develop a cost effective real-time foot pressure monitoring system that is capable of assisting identification of callus formation of diabetes patients. The device consists of a pressure sensing mat, a custom build data acquisition unit and the visualization software. For initial clinical testing of the system, normal subjects and subjects with diabetes who have calloused feet were considered.

\subsection{Existing Commercial Systems and Drawbacks}
Foot pressure mats \cite{5_abraham2011low} and foot insoles \cite{6_murphy2012foot} are the main methods of acquiring and analyzing the foot pressure. Pressure monitoring systems integrated to footwear have been used in \cite{7_amemiya2016establishment} and \cite{8_amemiya2013relationship}. However, according to results of \cite{9_zequera2010performance}, each subject requires a personalized insole. In \cite{10_al2015fabrication}, \cite{11_li2016design}, \cite{12_mattar2016low}, \cite{13_petsarb2012low} and \cite{14_razak2012design}, pressure mats and pads have been used to assess plantar pressure. The usage of pressure mats is not limited to the context of diabetes but also other applications including postural disorders \cite{12_mattar2016low} and post-surgical biomechanical assessment \cite{15_hahn2008changes}.

Tekscan Inc. and Novel gmbh are two leading pressure mat producers and their products, \cite{17_EMed} and \cite{18_Tekscan}, are widely used for commercial applications. However, the high cost and constrained data access limiting the ability to perform the required analysis and arriving at unique conclusions are the main drawbacks of these systems. 

\subsection{Proposed Solution}
The proposed solution is designed to overcome above mentioned drawbacks with a real-time pressure monitoring platform with access to raw pressure data which can be used on custom processing algorithms. 

The proposed device uses an integrated design to improve the sensor resolution from a standard off-the-shelf low resolution pressure mat, a high speed data acquisition system and a software for data visualization, post processing and analysis. The developed system is then used to assess foot pressure variations of normal subjects and subjects with calloused feet.

\section{Methodology}
The implementation of the project consisted of 3 sections; Design of the pressure mat, design of the data acquisition unit and the development of the user interface for data analysis.

\subsection{Design of the Pressure Mat}
\begin{figure}[htbp]
%
%
%
\centerline{\includegraphics[scale = 0.35]{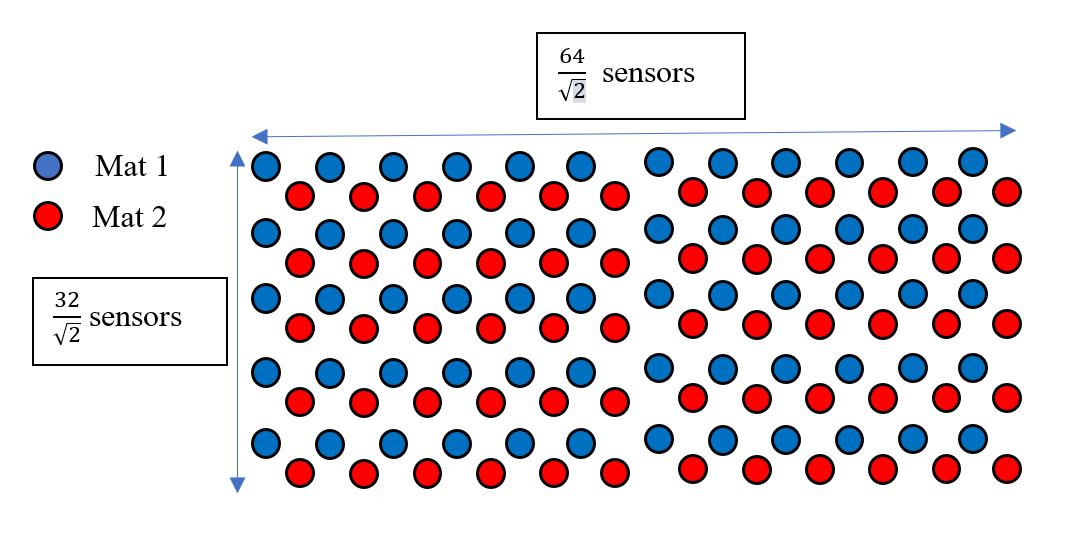}}
\caption{Sensor overlay with increased resolution from 1 to 2 sensors per cm$^2$. Placing a sensor among 4 sensors effectively contributes to $\sqrt[]{2}$ proportion along each dimension}
\label{fig1}
\end{figure}

\begin{figure}[htbp]
\centerline{\includegraphics[scale = 0.8]{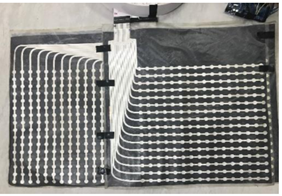}}
\caption{Composite pressure mat}
\label{fig2}
\end{figure}

The pressure mat was implemented by overlapping four pressure mats manufactured by Sensing-Tex, Barcelona, Spain. This was a standard product of Sensing-Tex thus the price was comparatively low (\texteuro 99). Each sensor mat consist of 256 (16x16) sensors with a sensor resolution of 1 sensor per cm$^2$ separately. We overlapped two such pressure mats to increase the number of sensors to 512 with a resolution of 2 sensors per cm$^2$. To increase the area covered, two such overlapped mats were placed next to each other, accordingly the final pressure mat included 1024 (32x64) sensors. The alignment of the pressure mats was done in such a way that a sensor of one mat lies among four sensors of the other mat symmetrically. This results in increased spatial resolution and independent data acquisition which increases the frame rate. The arrangement of the sensors shown in Fig.~\ref{fig1} and the final arrangement of the composite mat is shown in Fig.~\ref{fig2}.
%
%
%
%

\subsection{Data Acquisition System}
\begin{figure}[htbp]
\centerline{\includegraphics[scale = 0.28]{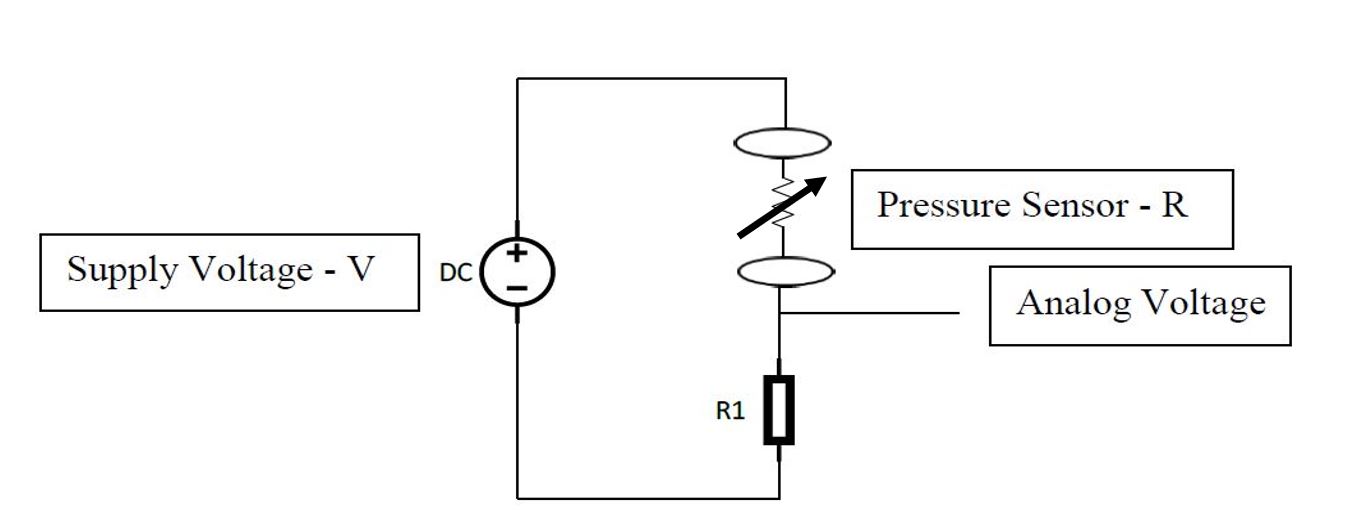}}
\caption{Potential divider configuration of the sensor interface circuit}
\label{fig3}
\end{figure}

Data was acquired to the system at a scanning rate of 155 Hz and the image was visualized in a MATLAB user interface. The pressure mat is a piezoresistive sensor array that contains a conductive ink that changes its resistance according to the force applied on it. Therefore, the sensor can be considered as the variable resistor and the analog data is acquired by the system through a potential divider as shown in Fig.~\ref{fig3}.

The analog voltage of the potential divider is given by \eqref{eq1},
\begin{equation}
Analog Voltage = \frac{R_1}{R_1+R_{sensor}}V_{supply}\label{eq1}
\end{equation}

\begin{figure}[htbp]
\centerline{\includegraphics[scale = 0.45]{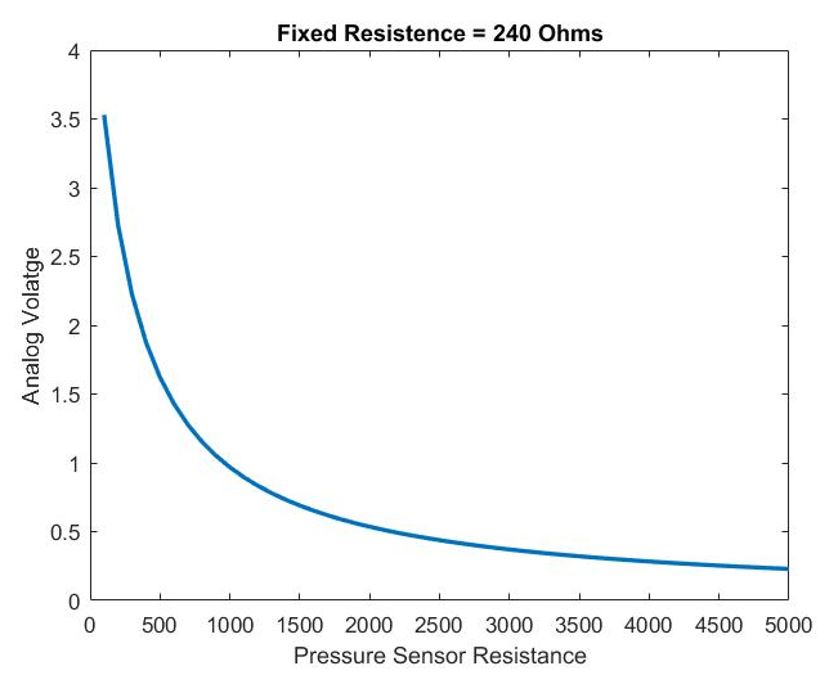}}
\caption{Pressure-sensor resistance curve}
\label{fig4}
\end{figure}

In the hardware design stage, the value for the fixed resistor \(R_1\) is pf paramount as the sensitivity of potential divider depends on it. The pressure sensor resistance graphs were obtained for a range of resistances out of which, 240\si{\ohm} was selected. The selection criteria was based on the output voltage factor and the sensitivity factor i.e. resistor with the value 240\si{\ohm} was selected for limiting the output voltage to 3.3V for over-voltage protection of the data acquisition system without affecting the sensitivity. The resistance curve of the pressure sensor against voltage from the potential divider is shown in Fig.~\ref{fig4}.

A printed circuit board (PCB) was designed to drive the sensors and acquire the analog data from the sensors. The sensors were driven by transistors via a decoder system as the output of decoders was not sufficient to drive each row of the sensor matrix. Data was acquired to the system through multiplexers via the potential divider as shown in Fig.~\ref{fig3}. STM32F3 Discovery development board with an ARM\textsuperscript{\textregistered} Cortex\textsuperscript{\textregistered} M4-based mixed signal MCU was used as the main controller which can be run with a real-time operating System. The board voltage level is 3.3 V and possesses a higher analog to digital conversion rate. However, other than the clock and processing capability, availability of development platform, smaller size, low cost and easier integration to circuitry, extendibility of system and adequate amount of digital and analog pins are the other factors considered in selecting the microprocessor. STM32F303VCT has four analog to digital converter blocks which operates at a maximum of 72 MHz and supports a maximum resolution of 12 bits. It consists of 39 analog pins and USB 1.1 full-speed up to 12 Mbits per second which satisfies the requirements of the proposed data acquisition system. 

\begin{figure}[htbp]
\centerline{\includegraphics[scale = 0.8]{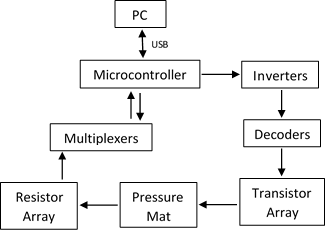}}
\caption{System architecture}
\label{fig5}
\end{figure}

During the troubleshooting process, it was found that the system does not work properly as a result of TTL-CMOS issue. Hence, the controlling signals were interfaced through an inverter circuit. The finalized hardware architecture is shown in Fig.~\ref{fig5}. Data was acquired to the PC via the USB interface. The programming of STM32F3 discovery was done using the Keil uVision 5 and STM32 Cube MX software.

\begin{figure}[htbp]
\centerline{\includegraphics[scale = 0.58]{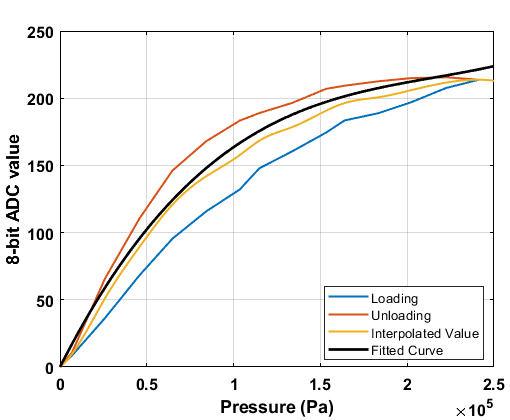}}
\caption{Calibration curve obtained by combining loading and unloading curves followed by interpolation and curve fitting}
\label{fig11}
\end{figure}

The system was calibrated using a weight baring base which covered 100 cm$^2$ of sensor area. The loading and unloading curves were obtained with a range of known pressure values as shown in Fig.~\ref{fig11}. The effect of hysteresis is apparent, hence, a mean curve was fitted to obtain the calibration curve. 

A software tool was developed for pressure system, which enables visualization and analysis of pressure distributions of the feet. The initial step of the software tool is to convert the raw pressure data into meaningful values using the calibration curve. The software tool is integrated with the functionality of real-time visualization of data, saving data for analysis, outputting the parameters related to pressure  and controlling the communication between the PC and the device.

\section{Data Analysis}
Despite the main objective being the development of a comprehensive pressure monitoring system, the use of such a system for assessing plantar pressure distribution during standing and walking, and the employability of such a system in clinical applications was needed to be verified. Hence, a clinical test concerning the pressure variation of normal feet and feet subjected to callus formation was performed.

\subsection{Selection of Test Subjects}
For the initial study, 10 subjects with no visible foot complications (i.e. due to wounds or poor-fitting shoes) and 5 subjects with diabetes who have developed callus in either of the foot was selected. In 3 subjects, the callus was predominantly in the $1^{st}$ and $2^{nd}$ metatarsal heads of either of the foot and in the $5^{th}$ metatarsal head in the other 2 subjects. The patients were adjudicated to have callus based on the opinion of a clinical expert and patient history.

10 young adults (age 24$\pm$1 years) who were considered as healthy subjects had a mean body mass of 68$\pm$8kg. The subjects with calloused feet belonged to the age category of 52$\pm$8 years with a body mass range of 75$\pm$10kg,

\subsection{Data Extraction and Analysis}
In the initial study, the emphasis was on assessing the pressure variation in normal foot and foot with callus and the implications on the posture. Hence, the subjects were asked to stand still on the pressure mat to capture the pressure map. Due to high frame rate, 50 frames were captured (approximately within one third of a second) and averaged linearly.

\begin{figure}[htbp]
\centerline{\includegraphics[scale = 0.45]{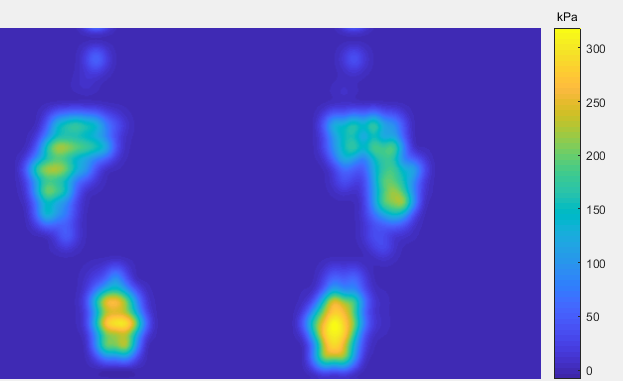}}
\caption{Processed pressure map of plantar}
\label{fig10}
\end{figure}

The raw data was then up-sampled 100-fold in spatial dimension (by a factor of 10 along x and y axes) and interpolated using 2-D spline interpolation followed by Gaussian smoothing according to \eqref{eq2} using a 5x5 kernel with a standard deviation of 0.8. The resulting pressure map obtained for a normal subject is shown in Fig.~\ref{fig10}.

\begin{equation}
G(x,y) = \frac{1}{2\pi\sigma^2}e^{-\frac{x^2+y^2}{2\sigma^2}}\label{eq2}
\end{equation}

Pressure maps obtained are then processed to extract important matrices related to posture and stresses related to callus formation. These are extensively described below.
\begin{itemize}
\item Mean foot pressure and center of pressure\\
Mean foot pressure (MFP) of left and right feet as given in \eqref{eq3} and the centers of pressure of each foot and the resultant center of pressure according to \eqref{eq4} were calculated for the generated pressure maps.

\begin{equation}
MFP=\frac{\sum_{n=1}^{N} P_n}{N}\label{eq3}
\end{equation}

\begin{equation}
X_{CP}=\frac{\sum_{n=1}^{N} P_n x_n}{\sum_{n=1}^{N} P_n}\label{eq4}
\end{equation}

\item Load distribution\\
The percentage of load applied on each foot was calculated by manual selection of the regions. Load was calculated by multiplying each pressure value by effective sensor area. Hence, when percentage is considered, effect of sensor area cancels out and is simplified to \eqref{eq5}.

\begin{equation}
Load_{left}=\frac{\sum_{n=1}^{N_L} P_n}{P_{total}}\label{eq5}
\end{equation}

\item Pressure on primary contact points\\
This section emphasizes on variation of pressure in primary contact points; metatarsal heads and heel which are also prominent areas of callus formation. These regions were annotated manually for analysis, and the mean pressure of these areas was compared between normal and calloused feet. This is also expressed as a percentage of load distribution as given in \eqref{eq6}.

\begin{equation}
Load_{left}=\frac{\sum_{}^{cal. region} P_n}{P_{total}}\label{eq6}
\end{equation}

\end{itemize}

\section{Results}
The results generated by the system could be analyzed under two categories.

\subsection{Subjects without foot complications}

The mean parameters calculated for normal subjects is summarized in the table \ref{table_MT_SD}.

\begin{table}[h]
\caption{Pressure parameters for normal subjects}
\label{table_MT_SD}
\begin{center}
\begin{tabular}{|p{3.0cm}|p{1.8cm}|p{1.8cm}|}\hline

\textbf{Parameters} & \textbf{Left} & \textbf{Right}\\\hline

{MFP} & 102.27 & 108.24\\\hline
{Load \%} & 46.01 & 53.99\\\hline
{Mean Heel Pressure} & 211.87kPa & 228.55kPa\\\hline
{Max Heel Pressure} & 282.11kPa & 308.49kPa \\\hline
{Load \% on Heel} & 39.13 & 43.46\\\hline
{Mean Metatarsal Pressure} & 188.24kPa & 187.87kPa\\\hline
{Max Metatarsal Pressure} & 223.89kPa & 215.62kPa\\\hline
{Load \% on Metatarsal} & 42.49 & 38.24\\\hline

\end{tabular}
\end{center}
\end{table}

\subsection{Comparison of normal subjects and calloused subjects}
Similar analysis was carried out for subjects with calloused feet with respect to diseased and normal foot as shown in table \ref{table_Callus}. However, the emphasis should be on how these values vary between the subject groups. It should be kept in mind that all the 5 subjects were selected so that all of them have calluses in metatarsal region of the foot.

\begin{table}[h]
\caption{Pressure parameters for calloused subjects}
\label{table_Callus}
\begin{center}
\begin{tabular}{|p{3.0cm}|p{1.8cm}|p{1.8cm}|}\hline

\textbf{Parameters} & \textbf{Calloused Foot} & \textbf{Normal Foot}\\\hline

{MFP} & 131.05 & 116.93\\\hline
{Load \%} & 55.42 & 44.58\\\hline
{Mean Heel Pressure} & 279.09kPa & 273.42kPa\\\hline
{Max Heel Pressure} & 340.72kPa & 297.85kPa \\\hline
{Load \% on Heel} & 60.28 & 46.59\\\hline
{Mean Metatarsal Pressure} & 249.37kPa & 183.76kPa\\\hline
{Max Metatarsal Pressure} & 313.94kPa & 209.80kPa\\\hline
{Load \% on Metatarsal} & 27.20 & 32.79\\\hline

\end{tabular}
\end{center}
\end{table}

\section{Discussion}
This study consists of two main aspects; development of a custom pressure mat with associated hardware for pressure monitoring and the assessment of plantar pressure in normal subjects and subjects with calloused feet. The pressure mat contained 1024 sensors and has a spatial resolution of 1 sensor per cm$^2$. The mat supported a data acquisition rate of 155 Hz which is comparable to commercial products \cite{17_EMed} and \cite{18_Tekscan}. Hence, supports real-time pressure monitoring as well as dynamic analysis despite the smaller size of the mat.

The results of the system for normal subjects indicated a slightly skewed load distribution (3\%) in normal people. This could be due to the inherent posture of the subjects while standing. There was minimal disparity in mean heel and metatarsal pressures in each foot which is approximately 17 kPa and 1 kPa respectively. According to these results, both the heel and metatarsal bears about 40\% of the load on each foot in normal standing position. The peak heel and metatarsal pressures were not drastically different across the two feet. 

As anticipated, the system gives several significantly different measurements when subjects with callus were assessed. The mean foot pressure of the calloused foot was shown to be higher compared to the normal foot. Even though the mean heel pressures were not different across two feet, the percentage of load of each foot acting on the heel is high in calloused feet. In comparison to the normal subjects, the heel pressures are high in subjects with calloused feet even though the callus is formed in the metatarsal. The noteworthy point is the excessive pressure distribution in the metatarsal which is about 60 kPa than the healthy foot and feet of normal subjects. The peak metatarsal pressure of calloused feet is 90 kPa greater than that of normal feet. This verifies the increase of pressure due to callus formation. 

The limitations of this system include the inadequacy of the surface area for dynamic analysis and the constraints of the sensor characteristics for the analysis of high pressures which are exerted especially during running. The spatial resolution of the sensor mat could be improved to match currently existing high resolution devices. Hence, this system can further be improved to perform dynamic analysis while walking and running, by upscaling the hardware platform. 

\section{Conclusion}
Plantar pressure monitoring systems with sufficient accuracy is able to detect variations of foot pressure due to formation of callus. This is a vital precursor of foot ulceration due to diabetes. Hence, such systems could be used to monitor such patients to initiate early clinical intervention. 
The hardware system for the pressure mat was developed from scratch, thus it was challenging to match industry standards, which was ultimately accomplished. The low-cost system developed by us could be utilized for aforementioned vital clinical applications. 
The capabilities of such devices stretch beyond clinical applications, especially in biomechanical analysis in sports, rehabilitation, post-injury assessment etc. Fast data acquisition platforms, such as the one we have developed, would be ideal for the aforementioned applications.

\bibliographystyle{plain}
\bibliography{ref}
\nocite{*}


\end{document}